\documentclass[twocolumn,english,prl,amssymb,aps,superscriptaddress,showpacs,twocolumn,amsmath,showkeys,floatfix]{revtex4-2}

\usepackage{geometry}
\usepackage[active]{srcltx}
\usepackage{textcomp}
\usepackage{amsmath}
\usepackage{graphicx}
\usepackage{color}
\usepackage{esint}

\makeatletter

\usepackage[english]{babel}
\usepackage{physics}
\usepackage[breaklinks]{hyperref}
\usepackage{inputenc}

\makeatother

\begin{document}

\title{Multimode resonance transition to collapsed snaking in normal dispersion Kerr resonators: Bright versus dark solitons}

\author{Yifan Sun}\email{yifan.sun@uniroma1.it}
\affiliation{Department of Information Engineering, Electronics and Telecommunications, Sapienza University of Rome, Via Eudossiana 18, 00184 Rome, Italy}
\author{Stefan Wabnitz}
\affiliation{Department of Information Engineering, Electronics and Telecommunications, Sapienza University of Rome, Via Eudossiana 18, 00184 Rome, Italy}
\author{Pedro Parra-Rivas}
\affiliation{Department of Information Engineering, Electronics and Telecommunications, Sapienza University of Rome, Via Eudossiana 18, 00184 Rome, Italy}
%\author{Carles Mili\'{a}n}
%\affiliation{Institut Universitari de Matem\`{a}tica Pura i Aplicada, Universitat Polit\`{e}cnica de Val\`{e}ncia, 46022 Val\`{e}ncia, Spain}
% \author{Mario Ferraro}
% \affiliation{Department of Information Engineering, Electronics and Telecommunications, Sapienza University of Rome, Via Eudossiana 18, 00184 Rome, Italy}
% \author{Fabio Mangini}
% \affiliation{Department of Information Engineering, Electronics and Telecommunications, Sapienza University of Rome, Via Eudossiana 18, 00184 Rome, Italy}
% \author{Mario Zitelli}
% \affiliation{Department of Information Engineering, Electronics and Telecommunications, Sapienza University of Rome, Via Eudossiana 18, 00184 Rome, Italy}
% \author{Raphael Jauberteau}
% \affiliation{Department of Information Engineering, Electronics and Telecommunications, Sapienza University of Rome, Via Eudossiana 18, 00184 Rome, Italy}
% \author{Francesco Rinaldo Talenti}
% \affiliation{Department of Information Engineering, Electronics and Telecommunications, Sapienza University of Rome, Via Eudossiana 18, 00184 Rome, Italy}

\begin{abstract}
% We analyze the stability and dynamics of Kerr dissipative solitons (DKS) in the presence of a parabolic potential. This potential stabilizes oscillatory and chaotic regimes, favoring the generation of static DKS. Furthermore, the potential induces the emergence of new dissipative structures, such as asymmetric breathers and chimera-like states. Based on a mode decomposition of the previous dynamics, we unveil the underlying modal interactions.  

%	We study the dynamics of Kerr dissipative solitons (DKS) in the presence of a parabolic potential in normal dispersion regime. This potential supports multimode resonances (MMR), favoring the generation of high-order bright DKS. By reducing the potential strength, bright DKS gradually evolves to dark DKS, leading to the transition of multimode bifurcation structure to collapse snaking bifurcation structure. This dynamics is summarized in a phase diagram. 
We study the dynamics of Kerr cavity solitons in the normal dispersion regime, in the presence of an intracavity phase modulation. The associated parabolic potential introduces multimode resonances, which promote the formation of high-order bright solitons. By gradually reducing the potential strength, bright solitons undergo a transition into dark solitons. We describe this process as a shift from a multimode resonance to a collapsed snaking bifurcation structure. This work offers a comprehensive overview of cavity dynamics and may provide a potential pathway to access multi-stable states by effectively varying the phase modulation.

\end{abstract}

\maketitle

Temporal dissipative Kerr solitons (DKS) \cite{wabnitz:93} have emerged as a significant research topic in the field of photonics over the past decade. In the frequency domain, DKS are associated with the generation and manipulation of coherent frequency combs \cite{herr_temporal_2014-1,PASQUAZI20181}. 
DKS have been effectively generated in passive ring Kerr resonators, by mainly using two platforms: microresonators \cite{kippenberg_microresonator-based_2011} and macroscopic fiber rings \cite{leo_temporal_2010}.
The formation of DKS relies on a delicate equilibrium that encompasses both the counter-balance between dispersion and nonlinearity, as well as the balance between cavity loss and external driving field.
The dynamics and stability of DKS have been the object of detailed analysis in a mean-field approximation. In this framework, the evolution of the field recirculating in a passive Kerr resonator is described by a driven and damped nonlinear Schr{\"o}dinger equation (NLSE) \cite{haelterman_dissipative_1992,coen_modeling_2013}, also known as Lugiato-Lefever equation (LLE).  % The DKS is obtained by a standard perturbation analysis of the NLSE soliton solution  \cite{wabnitz:93}.
With increasing pump intensity, the intracavity field exhibits a variety of instabilities, resulting in intricate spatiotemporal dynamics, that can manifest as periodic phenomena, such as breathers, and chaotic behaviors \cite{leo_temporal_2010,coillet_routes_2014,anderson_observations_2016,liu_characterization_2017,lucas_breathing_2017,chen_experimental_2018,bao_observation_2018}.

In Kerr resonators, it is well-known that in the anomalous dispersion regime, "bright" DKS manifest as light pulses superimposed on a relatively low-intensity continuous wave (CW) background, maintaining their shape and energy throughout as they circulate within the cavity. In this context, single-peak DKS solutions can be computed by applying standard perturbation analysis to the conservative soliton solution of the NLSE \cite{wabnitz:93}.
In contrast, in a normal dispersion cavity, "dark" DKS, characterized by localized intensity dips embedded within a relatively high-intensity background, can be formed \cite{xue2015mode}.
Recent studies have shown that the presence of higher-order perturbations can bring about a fundamental change in the occurrence of bright and dark DKS, under various dispersion driving conditions.
Notably, the effects of third- and fourth-order dispersion \cite{tlidi_high-order_2010,parra-rivas_coexistence_2017,talla_mbe_existence_2017,TOD_bright_exper_norm_dis}, the influence of the stimulated Raman effect \cite{parra-rivas_influence_2020}, and the interaction with frequency-dependent cavity losses \cite{huang2015mode} may result in the emergence of novel forms of localized states, or in the coexistence of bright and dark DKS. Furthermore, the occurrence of dark-bright DKS bound states have been observed by seeding two modes with dispersion of opposite signs \cite{Zhang2022}.

In addition, 
intracavity phase modulation [e.g., via an electro-optical modulator (EOM)], offers an additional degree of control over DKS dynamics and serves as a valuable tool for investigating synthetic dimensions \cite{Tusnin2020,Englebert2021}.
Our recent studies in this framework have shown that the parabolic potential introduced by the EOM can lead to soliton stabilization  in the anomalous regime \cite{Sun2022}, and the emergence of a host of new solutions, such as high-order DKS, chaoticons \cite{Sun2023} and 3D DKS \cite{Sun2022b}. 
%Recently, our studies of the dynamics of a new type of DKS in the presence of the parabolic potential which is introduced by the EOM, led to predicting soliton stabilization in the anomalous regime \cite{Sun2022}, and the emergence of a host of new solutions, such as high-order DKS, chaoticons \cite{Sun2023} and 3D DKS \cite{Sun2022b}.
% In this regime, the field exhibits temporal localization at the position where the attractive parabolic potential has a maximum. 
% In anomalous dispersion, attractive localization for the bright DKS occurs at the position of maximum phase modulation, namely parabolic potential has a maximum, while the position  of minimal phase modulation is repulsive.
In anomalous dispersion, DKS will always drift along an increasing phase gradient\cite{Erkintalo2022}. Therefore, attractive localization for bright DKS occurs at the maximum position of parabolic potential. 
% Conversely, the position of lower phase modulation exhibits repulsive behavior.

In this letter, we remarkably show that bright DKS can be generated even in a normal dispersion cavity with the presence of local minimal phase modulation. %can be achieved by inverting the sign of both anomalous dispersion and parabolic potential.
%Specifically, we observe that this bright DKS state occurs precisely when normal dispersion is present alongside a minimal parabolic potential position.
% In this letter, w
Our theoretical investigation demonstrates the significant influence of the parabolic potential in the transition between bright and dark DKS, which occurs within a dissipative Kerr resonator operating under normal dispersion.
The introduction of the parabolic potential facilitates the generation of multimode bright DKS.
By reducing the potential strength, the temporal duration of high-order bright DKS grows larger, eventually transitioning towards dark DKS. This phenomenology is comprehensively analyzed by using a systematic bifurcation analysis, which permits to establish a clear connection between the multimodal bifurcation structure and the collapsed homoclinic snaking, that appears in the absence of external phase modulation \cite{parra-rivas_origin_2016}.
%
%
%The parabolic potential approximates a periodic (e.g., sinusoidal) potential around the center of the DKS. Specifically, we find that, for low pump values, the potential stabilizes oscillatory and chaotic dynamics in favor of static DKS. As the pump power grows larger, the potential induces the appearance of asymmetric breathers and {\it chaoticons}, i.e. chimera-like states, where the background field state coexists with an incoherent spatio-temporal chaotic state. Moreover, chaoticons coexist with single-peak DKS, and form a hysteresis loop. To support our findings, we carry out a systematic bifurcation analysis, which establishes the connection with the multimodal structure of the potential. 
\begin{figure}[!t]
\centering
\includegraphics[scale=1]{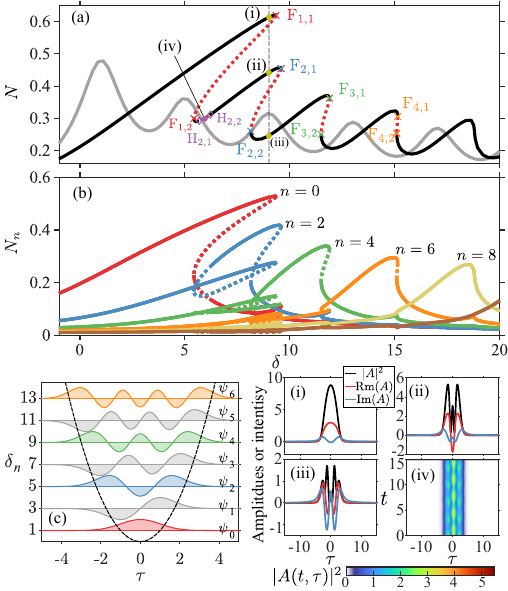}
\caption{(a) Bifurcation diagrams of stationary solutions showing the averaged pulse amplitude $N=\sqrt{E/T}$ vs. $\delta$ in the presence (black and red curves) and absence (gray curve) of Kerr nonlinearity, when $(P,C)=(2.5,-1)$. (b) Bifurcation diagram of the averaged mode components $N_n=\sqrt{C_n^2/T}$ vs. $\delta$. Stable (unstable) solutions are plotted by solid (dashed) curves. The FB pairs are marked by $F_{m,1}$ and $F_{m,2}$ at each resonance $m$. $H_{2,1}$ and $H_{2,2}$ are HB pairs. The DKS (i-iii) and breather (iv), marked in (a), are shown in subplots (i-iv), respectively. (c) Linear eigenmodes $\psi_n$, eigenvalues $\delta_n$ and potential $V(\tau)=-C\tau^2$.}
\label{fig1}
\end{figure}

In the mean-field approximation, the evolution of the light field in a coherently driven and phase-modulated passive cavity in the normal dispersion regime obeys the dimensionless equation \cite{Sun2022,Sun2023}
\begin{equation} 
\partial_t A = -\mathrm{i} \partial_\tau^2 A - \mathrm{i}C\tau^2A 
+ \mathrm{i}|A|^2A  - (1+\mathrm{i}\delta)A+ P,
\label{eq1}
\end{equation}
where $A$ is the slowly varying field envelope, $\tau$ and $t$ are fast and slow time, respectively. The term $-\partial^2_\tau A$ describes second-order normal dispersion, $-A$ and $\delta A$ represent linear loss and cavity phase detuning, $\mathrm{i}|A|^2A$ is the Kerr nonlinear term, and $P$ is the driving pump field amplitude.
The synchronous parabolic temporal potential $-\mathrm{i}C\tau^2$ from the EOM is introduced, where $C$ is the phase modulation curvature \cite{Sun2023}.
It is worth noting that the potential shape can be readily adjusted by designing the voltage profile incorporated into the EOM \cite{copie2023}. This flexibility allows for the implementation of various potential shapes, such as a parabola, linear, sinusoidal, or other desired profiles.
By changing the coordinates meaning, Eq.~(\ref{eq1}) also describes spatial bottle resonators \cite{Oreshnikov2017}.

To perform the bifurcation analysis of steady-state soliton solutions $A_s$ (i.e., $\partial_t A_{\rm s}=0$) of Eq.(1), we apply a combination of numerical techniques, including direct numerical simulations (DNS), path-continuation techniques through pde2path, and numerical linear stability analysis \cite{Uecker2014a}.
To illustrate the essential characteristics arising from the parabolic potential and Kerr nonlinearity, Fig.~\ref{fig1}(a)
compares the bifurcation structure of stationary solutions of Eq.(1) either in the absence or in the presence of the Kerr nonlinear term $\mathrm{i}|A|^2A$, as a function of detuning $\delta$, for the fixed 
$(P,C)=(2.5,-1)$. 
%where 
%the pulse norm $N\equiv$
%provides a comparison of bifurcation diagrams for stationary solutions under two scenarios: in the absence and presence of the Kerr nonlinear term $\mathrm{i}|A|^2A$, when varying the detuning $\delta$ but fixing the pump and the potential $(P,C)=(2.5,-1)$. 
Here, we plot the modification of the average field amplitude $N=\sqrt{E/T}$, with $E=\int|A|^2\dd \tau$, as a function of $\delta$ for a time domain window $T=100$. %denotes the temporal length along the fast time dimension.
The gray curve in Fig.~\ref{fig1}(a) shows the modification of $N$ in the absence of the Kerr nonlinearity. This curve consist of many peaks, equally spaced at $\delta=1,\,5,\,9,\,13,\,17,\cdots$, with a gap $\Delta\delta=4$. This pattern already illustrates the presence of multimode resonances (MMR), indicating that the cavity can sustain higher energy levels at these specific resonance points. 
% In fact, these resonances are exactly determined by eigenvalues $\delta_n$ and eigenmodes $\psi_n$ of the linear system operator $\hat{H_0}=[-\mathrm{i} \partial_\tau^2 - \mathrm{i}C\tau^2 ]$, which result from the interaction between second-order dispersion and parabolic potential. The modes obeys the linear eigenvalue equation $\delta_n\psi_n=\hat{H_0}\psi_n$, therefore, we obtain eigenvalues $\delta_n=2\sqrt{|C|}(n+1/2)$ and eigenmodes $\psi_n$ following Hermite-Gaussian (HG) functions $\psi_n(\tau) = (2^nn!)^{-\frac{1}{2}}\pi^{-\frac{1}{4}}\exp\left(-\frac{\tau^2}{2a_\tau^2}\right) H_n\left(\frac{\tau}{a_\tau}\right)$, where $H_n$ is the Hermite polynomial and $a_\tau=|C|^{-1/4}$ is the scaling ratio. 

These MMR find their precise determination in the eigenvalues $\delta_n$ and eigenmodes $\psi_n$ of the linear system operator $\hat{H}_0=[-\mathrm{i} \partial_\tau^2 - \mathrm{i}C\tau^2 ]$. These quantities emerge as a result of the intricate interplay between second-order dispersion and the parabolic potential. The modes obey the linear eigenvalue equation $\delta_n\psi_n=\hat{H}_0\psi_n$, enabling us to obtain the eigenvalues $\delta_n=2\sqrt{|C|}(n+1/2)$ and eigenmodes $\psi_n$, which are characterized by Hermite-Gaussian (HG) functions, namely %functions. The eigenmodes are mathematically expressed as 
$\psi_n(\tau) = (2^nn!)^{-\frac{1}{2}}\pi^{-\frac{1}{4}}\exp\left(-\frac{\tau^2}{2a_\tau^2}\right) H_n\left(\frac{\tau}{a_\tau}\right)$, where $H_n$ represents the Hermite polynomial, and $a_\tau=|C|^{-1/4}$ is the scaling ratio. 
The relation among eigenmodes $\psi_n(\tau)$, eigenvalues $\delta_n$, and their parabolic potential $\tau^2$ is illustrated in Fig.~\ref{fig1}(c). Homogeneous pumping effectively suppresses resonances of asymmetric HG modes, resulting in a resonance gap of $\Delta\delta=4\sqrt{|C|}$ between consecutive resonances \cite{Sun2023}. 
% Note that, in this work, we study the cases of the field locating at the minimal phase modulation point ($C<0$) in the normal dispersion regime ($-\mathrm{i} \partial_\tau^2A$), where the signs for dispersion and potential in Eq.~(\ref{eq1}) are opposite with respect to the model in Ref.~\cite{Sun2023}. As a result, the linear eigenvalue locates positive values, while eigenfunctions are the same profiles. 
Note that in this work, we focus on scenarios where the field is located at the minimal phase modulation point ($C<0$) within the normal dispersion regime ($-\mathrm{i} \partial_\tau^2A$). The signs of dispersion and potential in Eq.(\ref{eq1}) are opposite, when compared with the model presented in Ref.\cite{Sun2023}. As a consequence, the linear eigenvalues are positive, while the eigenfunctions exhibit the same profiles.

\begin{figure}[tp]
\centering
\includegraphics[scale=1]{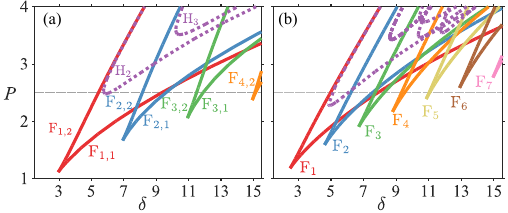}
\caption{$P$ vs. $\delta$ phase diagram showing FB-connected and HB-connected  curves, when $C=-1$ in (a) and $C=-0.25$ in (b).}
\label{fig2}
\end{figure}

% \begin{figure}[tp]
% \centering
% \begin{tikzpicture}
% \node[anchor=south west,inner sep=0,align=center] (image) at (0,0) 
% %		{\begin{minipage}{1\columnwidth}
% {\begin{minipage}{1\columnwidth}
% \includegraphics[scale=1]{figures/_C_1p00_nor_dis_phase_diagram_P_vs_delta_.png}
% %				\includegraphics[scale=1]{figures/_nor_dis_phase_diagram_C_vs_delta_2_1.png}
% \includegraphics[scale=1]{figures/_C_0p25_nor_dis_phase_diagram_P_vs_delta_.png}	
% %				\includegraphics[scale=1]{figures/_nor_dis_phase_diagram_C_vs_delta_2_2.png}
% \end{minipage}};
% %		{\begin{minipage}{0.49\columnwidth}
% %		\includegraphics[scale=1]{figures/_nor_dis_phase_diagram_C_vs_delta_2_1.png}
% %		\includegraphics[scale=1]{figures/_nor_dis_phase_diagram_C_vs_delta_2_2.png}
% %		\end{minipage}}
% \begin{scope}[x={(image.south east)},y={(image.north west)}]
% \draw (0.10,0.83)	 node{\normalsize  {\color{black}$\rm (a)$}};		
% %			\draw (0.10,0.43)	 node{\normalsize  {\color{black}$\rm (b)$}};		
% \draw (0.60,0.83)	 node{\normalsize  {\color{black}$\rm (b)$}};		
% %			\draw (0.64,0.43)	 node{\normalsize  {\color{black}$\rm (d)$}};	
% \draw [dashed] (0.07,0.555) -- (0.44,0.555);			
% \draw [dashed] (0.567,0.555) -- (0.95,0.555);
% \end{scope}
% \end{tikzpicture}
% \caption{$P$ vs. $\delta$ phase diagram for FB-connected curves, when $C=-1$ in (a) and $C=-0.25$ in (b). {\color{red} (Adding Fold point markers.)}
% }	
% \label{fig2}
% \end{figure}

\begin{figure*}[!t]
\centering
\includegraphics[scale=1]{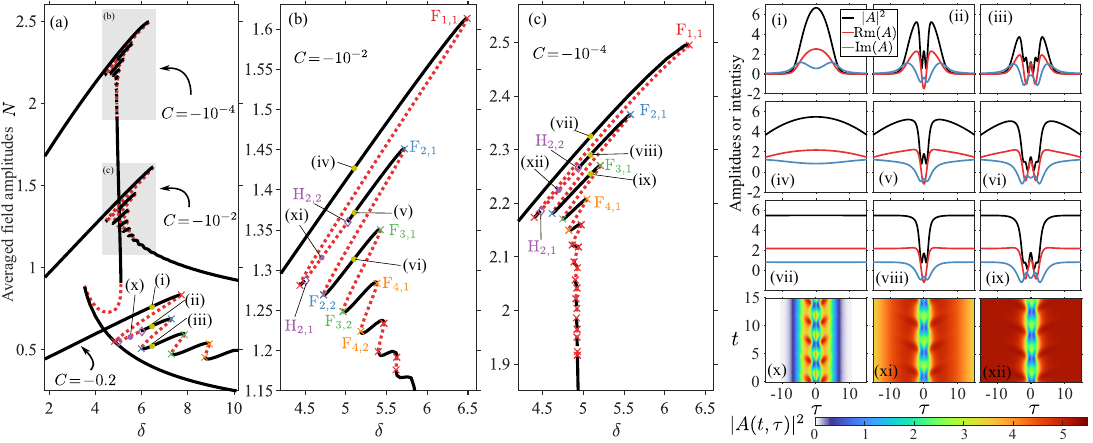}
\caption{
Bifurcation diagram showing the averaged pulse amplitude $N=\sqrt{E/T}$ vs. $\delta$, when $C=-0.2$, $C=-\times10^{-2}$ and $C=-10^{-4}$ in (a). The zoomed regions for $C=-\times10^{-2}$ and $C=-10^{-4}$ are plotted in (b) and (c), respectively. The individual DKS and breathers marked by points (i-xii) in (a,b,c) are plotted in (i-xii). The pump $P=2.5$.
}
\label{fig3}
\end{figure*}

With the inclusion of the Kerr effect, noticeable changes occur in the resonance peaks [see the black curve in Fig.~\ref{fig1}(a)]. This is because the Kerr effect introduces an intensity-dependent phase modulation, thus altering the effective detuning as $\mathrm{i}(|A|^2-\delta)$.
For resonance peaks with higher intensity $|A|^2$, a larger detuning $\delta$ is preferred, in order to compensate for the phase modification caused by $|A|^2$, resulting in the tilting of the resonance peaks. These tilts lead to overlapping portions of the solutions on the right side of the resonance peaks. Consequently, fold bifurcations (FB) arise (see $\mathrm{F}_{n,1}$ and $\mathrm{F}_{n,2}$ marked at resonance peak $n$), leading to the emergence of branches of multi-stable states. The solid (dashed) curves in the bifurcation diagram represent stable (unstable) solutions, which have been verified by a linear stability analysis.

Figures~\ref{fig1}(i,ii,iii) show three stable solutions for the same value of detuning $\delta=4.9$. Each solution exhibits a different number of peaks: 1, 3, and 5, respectively. These differences originate from the dominance of a Hermite-Gaussian (HG) mode with index $n=0,2,4$.
To verify this, we may project the field onto the HG mode basis [see Fig.~\ref{fig1}(b)]. This allow for determining the mode amplitudes $C_n(t) = e^{-i\delta_nt} \int_{-\infty}^{\infty}A(\tau,t)\psi_n(\tau)\dd \tau$, which are represented in Fig.~\ref{fig1}(b) in terms of the averaged mode amplitude $N_n=\sqrt{C_n^2/T}$.
%different $\delta$. 
% As $\delta$ increases, the mode component $n-1$ increases significantly when $\delta$ approaches the corresponding resonance peak $n$, indicating the resonance is dominant by the coresponding mode.
As $\delta$ increases, the component of mode $2(n-1)$ exhibits a significant growth, particularly when $\delta$ approaches the corresponding resonance peak $n$. This indicates that each resonance is predominantly associated with a given mode.
Furthermore, we observe the existence of breather solutions within the two Hopf bifurcations (HB) $\mathrm{H}_{2,1}$ and $\mathrm{H}_{2,2}$ which occur at the second resonance peak [see Fig.~\ref{fig1}(a)].

%---------------------------------------------

By increasing the pump strength, more energy gets coupled into the cavity, leading to potentially richer dynamics. To explore the impact of pump strength $P$ on the system behavior, we constructed phase diagrams as in Fig.~\ref{fig2}, by varying both $P$ and the detuning $\delta$. In Fig.~\ref{fig2}(a), the FB-point-connected curves $\mathrm{F}_n$ for each resonance peak $n$ are displayed, for $C=-1$. It is observed that increasing the pump $P$ results in larger tilts for all resonance peaks, and a higher number of occurrences of FB at higher-order resonances. Additionally, larger regions for breathers are observed as the pump is increased (see the dashed purple curves representing the HB-point-connected curves $\mathrm{H}_2$ and $\mathrm{H}_3$ at resonances $n=2,3$). On the other hand, decreasing the pump strength eliminates multiple stable and unstable states, leading the cavity system to gradually return to its linear state.

The strength of the potential has a significant impact on the cavity dynamics: not only the potential determines the eigenvalue $\delta_n$, but it also dictates the scaling factor $a_\tau = C^{-1/4}$ for the size of the HG modes. This implies that reducing the potential strength leads to a denser distribution of resonance points in $\tau$, while resulting in a wider field distribution. To examine this effect, we have calculated the phase diagram with a potential strength of $C=-0.25$, corresponding to the linear resonance separation $\Delta\delta=2$. As depicted in Fig.~\ref{fig2}(b), the phase diagram exhibits similar dynamics to Fig.~\ref{fig2}(a), but with a greater number of FB (solid curves) and HB (dashed purple curves), occurring at higher-order resonances within the same region. 
%This confirms that decreasing the potential strength leads to an increased density of high-order resonances in the system dynamics.

At this point, a natural question arises, regarding the bifurcation structures and DKS profiles when the potential strength $C$ is reduced: do these high-order resonances vanish, and how do high-order DKS modify? To address this question, in Fig.~\ref{fig3}(a) we plot the bifurcation diagrams for progressively reduced potential strengths, namely $C=-0.2,\, -10^{-2},\,-10^{-4}$, corresponding to the eigenvalue differences $\Delta\delta=0.89,0.2,0.02$, respectively.

\begin{figure}[tp]
\centering
\includegraphics[scale=1]{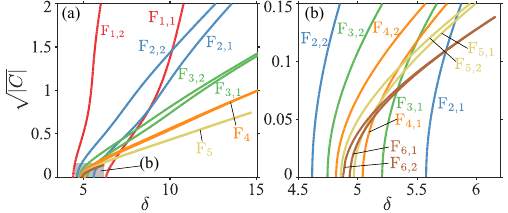}
\caption{$\sqrt{|C|}$ vs. $\delta$ phase diagrams showing FB-point-connected curves $\mathrm{F}_{n,1}$ and $\mathrm{F}_{n,2}$ for each resonance peak $n$ in (a) and a zoomed region in (b), when $P=2.5$.}
\label{fig4}
\end{figure}

For $C=-0.2$, the bifurcation diagram exhibits a similar pattern to Fig.~\ref{fig1}(a), but within a narrower detuning region. 
Three individual solutions with $\delta = 6.5$, marked (i-iii) in Fig.~\ref{fig3}(a), are plotted in Fig.~\ref{fig3}(i-iii). When compared with the solutions in Fig.~\ref{fig1}(i-iii), it is evident that the solutions in Fig.~\ref{fig3}(i-iii) have wider temporal extension. For the reduced strength $C=10^{-2}$, the close-up view of Fig.~\ref{fig3}(a) shown in Fig.~\ref{fig3}(b) exhibits a striking resemblance in the bifurcation structure to the previous scenarios. Now the linear eigenvalue separation becomes very small $(\Delta\delta=\times10^{-2})$, while the mode scale is very large $a_\tau=C^{-1/4}=3.16$. 
As a result, the full width at half-maximum of the fundamental mode increases to $1.665a_\tau=5.27$, and the high-order modes become much wider. Figures~\ref{fig3}(iv-vi) showcase examples of DKS states at various branches, specifically for $\delta=5.1$. Note that, these states seem dark DKS emerging within a non-uniform domain, and may be similar to those arising in pulse-pumped cavities \cite{garbin_experimental_2017}.

%Therefore, the solutions (iv-vi) with $\delta=5.1$ exhibit a much longer temporal duration, indicating a transition towards dark DKS.
By further reducing $C$, the bifurcation structure in Fig.~\ref{fig3}(b) tends to rotate to the right leading, eventually, to the {\it collapsed snaking} shown in  Fig.~\ref{fig3}(a),(c) for to $C=-10^{-4}$ \cite{parra-rivas_origin_2016}. This structure consists of a sequence of DKS state branches that oscillate back and forth, in a damped fashion, around the Maxwell point of the system ($\delta=4.85$). Around this point, dark DKS, like those depicted in Fig.~\ref{fig3}(vii, viii, ix) for $\delta=5.1$ originate due to uniform front locking \cite{parra-rivas_origin_2016,Schelte:19}.  

%down to $C=-10^{-4}$, it becomes evident that all resonances converge towards a specific value of $\delta=4.85$ in the bifurcation diagram shown in Fig.~\ref{fig3}(c) (or Fig.~\ref{fig3}(a)). 

%This behavior indicates a convergence towards a collapse bifurcation structure, which was previously studied \cite{pedro_snakes_2018}. 
%The corresponding individual solutions with $\delta=5.1$ are depicted in Fig.~\ref{fig3}(vii, viii, ix), exhibiting the formation of dark DKS profiles.
%This type of solution was also observed experimentally, albeit using a pulsed pump scheme \cite{Garbin2017}.

Looking at it from another perspective, when increasing the potential strength, 
% the uniform edge background related to the dark DKS formation are deformed such that the latter smoothly become bright DKS.
the uniform background field on the two sides of the dark DKS undergoes deformation, resulting in a smooth transition to bright DKS. 
% This confinement occurs at the minimum phase modulation point in the potential within the normal dispersion regime, in contrast to the maximum phase modulation point within for anomalous dispersion \cite{Sun2022,Sun2023}. 
Such potential-induced confinement also applies to breather solutions. Breathers, marked by (x, xi, xii) in Fig.~\ref{fig3}(a, b, c), are plotted in Fig.~\ref{fig3} (x, xi, xii), showcasing the impact of the potential on their formation. 
In the bifurcation diagram, the potential effectively separates all multi-stable state branches in a collapse snaking structure towards the high detuning regime, leading to an MMR structure. 

The transition dynamics between high and low potential strength can be effectively visualized through the $\sqrt{|C|}$ vs. $\delta$ phase diagrams in Fig.~\ref{fig4}. In Fig.~\ref{fig4}(a), we plot the FB-point-connected curves $\mathrm{F}_{n}$ for each resonance peak $n$, while varying $\sqrt{|C|}$ and $\delta$. A zoomed region is shown in Fig.~\ref{fig4}(b). 
% As the potential strength increases, the eigenvalues (or resonance peaks) exhibit almost linear increments with wider gaps.
Increasing potential strength $\sqrt{C}$ weakens the influence of Kerr nonlinearity, reducing the resonance tilts and causing the gradual disappearance of the FB-point-connected curves $\mathrm{F}_{n}$ [see $F_5$ in \ref{fig4}(a) and $F_6$ in \ref{fig4}(b)].
Conversely, reducing $\sqrt{C}$ causes the resonances to converge towards the Maxwell point $\delta=4.85$. Notably, lower potential strengths give rise to the appearance of additional FB points (only 6 FB curves are plotted here). Further decreasing $\sqrt{C}$, higher-order resonances undergo the evolution that ultimately leads to the emergence of a collapse-snaking structure.

In this study, we utilized bifurcation analysis to uncover the transition from bright to dark DKS in a driven passive nonlinear cavity in the normal dispersion regime with a parabolic potential. Our analytical study of the linear eigenmodes demonstrated how the parabolic potential localizes the field even in the presence of normal cavity dispersion, leading to the emergence of high-order modes and resonances in the system. These resonances, affected by Kerr nonlinearity, lead to the formation of high-order bright multimode DKS and breathers. Furthermore, by reducing the potential strength, the DKS bifurcation structure shows denser resonances and an increased temporal duration of bright DKS. Eventually, the former converges to the well-known collapse snaking structure, and the latter modify into dark DKS. Our study not only provides physical insight into the nonlinear dynamics modification from MMR to collapse snaking, but also offers a pathway to realize multi-stable states in such cavity systems.

% \begin{backmatter}
\section*{Acknowledgements}
This work was supported by EU under the NRRP of NextGenerationEU, partnership on “Telecommunications
of the Future” (PE00000001 - “RESTART”), Marie Sklodowska-Curie Actions (101023717,101064614), Sapienza University of Rome Additional Activity for MSCA (EFFILOCKER).

\bibliography{main}

\end{document}